\documentclass[elsarticle,nofootinbib,twocolumn,superscriptaddress]{revtex4}

\usepackage{graphicx}
\usepackage{amsmath}
\usepackage{slashed}

\newcommand{\be}{\begin{equation}}
\newcommand{\ee}{\end{equation}}
\newcommand{\br}{\begin{eqnarray}}
\newcommand{\bea}{\begin{eqnarray}}
\newcommand{\eea}{\end{eqnarray}}
\newcommand{\er}{\end{eqnarray}}
\newcommand{\ba}{\begin{array}}
\newcommand{\ea}{\end{array}}
\newcommand{\bi}{\begin{itemize}}
\newcommand{\ei}{\end{itemize}}
\newcommand{\bn}{\begin{enumerate}}
\newcommand{\en}{\end{enumerate}}
\newcommand{\bc}{\begin{center}}
\newcommand{\ec}{\end{center}}

\newcommand{\beq}{\begin{equation}}
\newcommand{\eeq}{\end{equation}}

\newcommand{\gsim}{\lower.7ex\hbox{$\;\stackrel{\textstyle>}{\sim}\;$}}
\newcommand{\lsim}{\lower.7ex\hbox{$\;\stackrel{\textstyle<}{\sim}\;$}}

\def\mysection#1{{\bf #1.} }

\begin{document}

\title{Twin Peak Higgs}

\author{Matti Heikinheimo}
\affiliation{National Institute of Chemical Physics and Biophysics, R\"avala 10, 10143 Tallinn, Estonia}

\author{Antonio Racioppi}
\affiliation{National Institute of Chemical Physics and Biophysics, R\"avala 10, 10143 Tallinn, Estonia}

\author{Martti Raidal}
\affiliation{National Institute of Chemical Physics and Biophysics, R\"avala 10, 10143 Tallinn, Estonia}
\affiliation{Institute of Physics, University of Tartu, Estonia}

\author{Christian Spethmann}
\affiliation{National Institute of Chemical Physics and Biophysics, R\"avala 10, 10143 Tallinn, Estonia}


\date{\today}

\begin{abstract}

A broad class of models in which electroweak symmetry breaking originates from dynamics in a singlet dark sector, and is transferred to the Standard Model via the
Higgs portal, predicts in general strongly suppressed Higgs boson mixing with a singlet scalar. In this work we point out that at present this class of models
allows for the second phenomenologically acceptable solution with almost maximal mixing between the Higgs and the scalar singlet. This scenario predicts an
almost degenerate twin peak Higgs signal which is presently indistinguishable from a single peak, due to the limited LHC mass resolution. Because of that, the LHC
experiments measure {\it inclusive} Higgs rates that all must exactly agree with Standard Model predictions due to sum rules. We show that if the dark sector and Standard Model communicate only via the singlet messenger scalar that mixes with the Higgs, the spin independent direct detection cross section of dark matter is suppressed by the scalar mass degeneracy,  explaining its non-observation so far.
\end{abstract}

\maketitle

\section{Introduction}

After the discovery of the Higgs boson~\cite{Aad:2012tfa,Chatrchyan:2012ufa}, the Standard Model (SM) has been experimentally verified within $\mathcal{O}(20\%)$ accuracy \cite{Giardino:2013bma,Ellis:2013lra,Djouadi:2013qya,Falkowski:2013dza}. Yet the question of what sets the electroweak scale at ${\cal O}(100)$~GeV
remains unanswered. Furthermore, we know that there must be physics beyond the SM to explain the existence of dark matter (DM)~\cite{Ade:2013lta}.
If DM is a weakly interacting particle, its thermal relic abundance suggests a very similar mass scale~\cite{Bertone:2004pz}. 
It is natural to assume that the two scales have actually a common origin.

Motivated by this possibility, several groups have recently studied a generic class of models where a SM-singlet dark sector, with non-trivial internal dynamics, generates a scale that is transmitted to the classically scale invariant SM via the Higgs portal. Some of the dark sector particles are stable and communicate with the SM only via the portal coupling, hence making them a perfect DM candidate that can evade all existing experimental bounds on direct detection. Examples of such models are dark Coleman-Weinberg models \cite{Hempfling:1996ht}, dark Technicolor \cite{Hur:2011sv,Heikinheimo:2013fta} and dark supersymmetry \cite{Heikinheimo:2013xua}. The internal dynamics of the dark sector and the nature of the DM candidates can be very different in each particular model, but from the SM point of view they all have one general, model independent feature: the SM Higgs doublet $H$ couples to a singlet scalar mediator particle $s$ via the portal coupling
\begin{equation}
\lambda_{sH}|H|^2s^2.
\label{portal_coupling}
\end{equation}
The messenger field acquires a vacuum expectation value from the dynamics of the dark sector, which is transmitted to the electroweak sector via (\ref{portal_coupling}). Electroweak symmetry breaking and fermion masses are then generated with the usual Higgs mechanism, but the origin of the vacuum expectation value of the Higgs field is in the dynamics of the dark sector. When combined with classical scale invariance and with additional assumptions on the nature of high energy physics, this setup opens new possibilities to address 
the naturalness of the electroweak scale, as discussed in  \cite{Heikinheimo:2013fta,Farina:2013mla,Dubovsky:2013ira}.
Typically the portal coupling  (\ref{portal_coupling}) is small and the messenger field is heavy, suppressing the Higgs-singlet mixing, 
and thus the dark sector is almost decoupled from the SM. 
In this case it can only be observed via precision measurements of the Higgs couplings, which are likely beyond the accuracy reach of the LHC experiments.

In this letter we study this general setup of dark sector portal models, focusing on the possibility that the messenger field has observable couplings to the SM particles via strong mixing with the Higgs, and the masses of the two physical mass eigenstates are nearly degenerate, so that the observed Higgs signal is in fact the sum of two overlapping signals from the two particles. We show that despite of the large (maximal) coupling of new physics to the SM in this scenario, it has remained unobservable.
A dedicated analysis for a similar case was recently published by the CMS collaboration \cite{CMS2Higgs}.
The scenario of two or more Higgs bosons in the vicinity of 126 GeV has previously been studied in the context of two Higgs doublet models and supersymmetry \cite{Gunion:2012gc}. The main difference here is that in our model the Higgs is mixed with the messenger scalar, which is a SM singlet instead of an electroweak doublet. 
The crucial difference between those models and our scenario is that no deviations from the SM Higgs rates can be observed in our case if one studies only inclusive rates\footnote{Here we assume that the dark sector does not contain particles that are below the Higgs decay threshold $2m_\chi=m_H$, so that the Higgs can not decay into dark sector particles via the mixing. If such light particles exist in the dark sector, there will be additional constraints from invisible decay width of the Higgs.}. The CMS search \cite{CMS2Higgs} indicates that this is indeed the case for nearly maximal mixing. We also show that  the mass degeneracy of the 
Higgs boson with the singlet has another consequence: It suppresses the spin independent DM direct detection cross section regardless of the nature of the DM.
Thus this scenario can be resolved with higher mass resolution in the 14 ~TeV LHC and in future lepton colliders, and by discovering Higgs-mediated 
DM nuclear recoils in new generation dark matter direct detection experiments. For earlier work on Higgs mixing with scalars see {\it e.g.}~\cite{ScalarMixing} and references therein.

The structure of this paper is the following: We will introduce the effective model and describe the mixing phenomena in the next section, and study the phenomenology of dark matter detection in section \ref{DMsection}. We will conclude in the last section.

\section{The generic model and Higgs phenomenology}

We start with a generic potential for the messenger scalar $s$ and the scalar part $h$ of the SM Higgs,
\begin{equation}
V=\lambda_s s^4+\lambda_h h^4-\lambda_{sh}s^2h^2-\mu^2s^2,
\label{scalar_potential}
\end{equation}
where all the couplings $\lambda_i$ are assumed to be positive and the sign of the portal term is chosen to be negative to induce a negative $\mu$-term for the Higgs. Notice that here we write the potential in terms of the scalar part of the Higgs field and the portal coupling is normalized accordingly, $\lambda_{sh} = \frac12 \lambda_{sH}$, where $\lambda_{sH}$ is the coupling constant appearing in (\ref{portal_coupling}). The $\mu$-term of the messenger field is assumed to originate from the dark sector, the structure of which is here left unspecified. The potential is bound from below if
\begin{equation}
4\lambda_s\lambda_h > \lambda_{sh}^2,
\end{equation}
which is satisfied for natural values of the couplings, since we assume the portal coupling $\lambda_{sh}$ to be small.

The minimum of the potential is given by
\begin{equation}
s= v_s = \sqrt{\frac{2\lambda_h}{{4\lambda_h\lambda_s-\lambda_{sh}^2}}}\mu,
\label{svev}
\end{equation}
\begin{equation}
h=v_h = \sqrt{\frac{\lambda_{sh}}{2\lambda_h}}v_s.
\label{hvev}
\end{equation}
Note that, depending on the structure of the dark sector, there could be other dimensionful couplings beside the $\mu$-term of $s$ in the potential (\ref{scalar_potential}), such as $\Lambda s^3$ or $\Lambda^3 s$, but these will only affect the value of $v_s$ given in equation (\ref{svev}). The vacuum expectation value of $h$ will allways be given by equation (\ref{hvev}), independent of the structure of the dark sector, if dimensionful couplings in the SM-sector are forbidden due to classical scale invariance. Thus we can effectively describe the mixing-phenomena of any portal-model with the potential (\ref{scalar_potential}), treating $v_s$ or $\mu$ as a free parameter.

The square-mass matrix is given by
\begin{equation}
M = \begin{pmatrix} 8\lambda_s & -2\sqrt{2}\lambda_{sh}\sqrt{\frac{\lambda_{sh}}{\lambda_h}} \\ -2\sqrt{2}\lambda_{sh}\sqrt{\frac{\lambda_{sh}}{\lambda_h}} & 4\lambda_{sh} \end{pmatrix} v_s^2 = \begin{pmatrix} \mu_s^2 & \Delta_\mu^2 \\ \Delta_\mu^2 & \mu_h^2 \end{pmatrix}.
\label{squaremassmatrix}
\end{equation}
The eigenvalues of this matrix are the squared masses of the physical scalars
\begin{equation}
m_{S_{1,2}}^2=\frac12\left(\mu_s^2+\mu_h^2\pm\sqrt{(\mu_s^2-\mu_h^2)^2+4\Delta_\mu^4}\right),
\label{eigenvalues}
\end{equation}
and the mixing angle is given by
\begin{equation}
\tan(2\theta)=\frac{2\Delta_\mu^2}{\mu_s^2-\mu_h^2}. \label{tan2theta}
\end{equation}

We are interested in the case where the mass-eigenvalues (\ref{eigenvalues}) are almost degenerate, implying
\begin{equation}
 \mu_s^2\approx \mu_h^2,\,\,\,\, \Delta_\mu^2 \ll \mu_s^2,\mu_h^2.
\end{equation}
In terms of the couplings $\lambda_i$ this is achieved if the following parameters are small:
\begin{eqnarray}
\epsilon &=& \lambda_s-\frac12\lambda_{sh} \ll 1, \nonumber \\
\epsilon' &=& \sqrt{\frac{\lambda_{sh}}{2\lambda_h}} \ll 1.
\label{epsilon_definitions}
\end{eqnarray}
In terms of these parameters the mass eigenvalues and the mixing angle are
\begin{equation}
m_{S_{1,2}}^2=4v_s^2\left(\lambda_{sh}+\epsilon\pm\sqrt{\epsilon^2+\lambda_{sh}^2\epsilon'^2}\right),
\end{equation}
\begin{equation}
\tan(2\theta)=-\lambda_{sh}\frac{\epsilon'}{\epsilon}.
\end{equation}
From these expressions it is obvious that the mass splitting of the two states is small if $\epsilon, \epsilon' \ll 1$, and that the mixing is large if $\epsilon \ll \lambda_{sh}\epsilon'$.

Concretely, we want the masses to be close to the experimentally observed Higgs mass, $m_{S_{1,2}}^2= m_H^2\pm\delta_m^2$, where $m_H\approx 125$ GeV, implying
\begin{equation}
4v_s^2(\lambda_{sh}+\epsilon)=m_H^2,
\label{fix_mh}
\end{equation}
\begin{equation}
4v_s^2\sqrt{\epsilon^2+\lambda_{sh}^2\epsilon'^2}=\delta_m^2.
\label{fix_deltam}
\end{equation}

In order to reproduce the electroweak gauge boson masses we need to fix $v_h=v_{EW}=246$ GeV. This condition also acts as a sum rule between the experimental signals from the production of the two scalars, so that the combined signal will reproduce the SM cross section.
Equations (\ref{hvev}) and (\ref{epsilon_definitions}) give $v_h=\epsilon'v_s$, and we can solve for $v_s$ using (\ref{fix_mh}), yielding
\begin{equation}
v_s=\frac{m_H}{2\sqrt{\lambda_{sh}+\epsilon}}.
\label{vs}
\end{equation}
Thus
\begin{equation}
\epsilon'=\frac{v_h}{v_s}=2x\sqrt{\lambda_{sh}+\epsilon}
\label{epsilon_prime}
\end{equation}
and
\begin{equation}
\lambda_h=\frac{\lambda_{sh}}{8x^2(\lambda_{sh}+\epsilon)},
\end{equation}
where we have defined $x=v_h/m_H\approx 1.968$. Then our condition for large mixing, $\epsilon \ll \lambda_{sh}\epsilon'$ implies
\begin{equation}
\epsilon \ll \lambda_{sh}^\frac32.
\label{maximal_mixing}
\end{equation}
Using (\ref{fix_deltam}), (\ref{vs}) and (\ref{epsilon_prime}) we can now solve for $\epsilon$:
\begin{equation}
\epsilon=\lambda_{sh}\frac{\delta^4-2x^2\lambda_{sh}\pm\sqrt{\delta^4-4x^2\lambda_{sh}(1-x^2\lambda_{sh})}}{1-\delta^4},
\label{epsilon}
\end{equation}
where $\delta=\delta_m/m_H$. For this equation to have real solutions, the expression in the square root must be positive, which happens when
\begin{equation}
\lambda_{sh}>\frac{1+\sqrt{1-\delta^4}}{2x^2}\,\,\,\,\,\,{\rm or}\,\,\,\,\,\lambda_{sh}<\frac{1-\sqrt{1-\delta^4}}{2x^2}.
\label{lambda_sh_solutions}
\end{equation}
The first solution corresponds to a non-stable vacuum with $\lambda_s<0$, so we will restrict the discussion to the second case. For maximal mixing we need $\epsilon$ to be small, which happens when $\lambda_{sh}$ is close to the upper bound of equation (\ref{lambda_sh_solutions}).

We will now examine a few numerical values for the above solutions. First we fix $\delta=0.1$, which fixes the masses of the physical states to $m_{S_1}=125.62$ GeV, $m_{S_2}=124.37$ GeV. Then we take a couple of values for $\lambda_{sh}$, some close to the upper bound (\ref{lambda_sh_solutions}) to get maximal mixing, some further away. The rest of the couplings and parameters are then given by the above equations. The results are given in table \ref{results_table}, where we have defined $r_\epsilon=|\epsilon|/\lambda_{sh}$ as a measure of the fine-tuning between the parameters $\lambda_{sh}$ and $\lambda_s$. The couplings $\lambda_{s'}$, $\lambda_{h'}$ and $\lambda_{s'h'}$ are the self-couplings of the physical scalar states $s'$ and $h'$ after rotating the fields $s$ and $h$ to the mass-eigenbasis.

\begin{table*}
\begin{center}
\begin{tabular}{|c|c|c|c|c|c|}
 \hline
 Point & $\lambda_{sh}$ & $\lambda_s$ & $\lambda_h$ & $r_\epsilon$ & $\mu$ \\ \hline
 A & $6.44845\times 10^{-6}$ & $3.22248\times 10^{-6}$ & $3.2283\times 10^{-2}$ & $2.7\times 10^{-4}$ & 62.49 GeV \\
 B & $6.39036\times 10^{-6}$ & $3.1891\times 10^{-6}$ & $3.2305\times 10^{-2}$ & $9.5\times 10^{-4}$ & 62.47 GeV \\
 C &  $5.80941\times 10^{-6}$ & $2.88665\times 10^{-6}$ & $3.2375\times 10^{-2}$ & $3.1\times 10^{-3}$ & 62.40 GeV \\
 D &  $6.4549\times 10^{-7}$ & $3.1669\times 10^{-7}$ & $3.2581\times 10^{-2}$ & $9.4\times 10^{-3}$ & 62.20 GeV \\
 E &  $6.4549\times 10^{-8}$ & $3.1639\times 10^{-8}$ & $3.2596\times 10^{-2}$ & $9.9\times 10^{-3}$ & 62.19 GeV \\
 F &  $3.2275\times 10^{-9}$ & $1.5818\times 10^{-9}$ & $3.2597\times 10^{-2}$ & $9.9\times 10^{-3}$ & 62.19 GeV \\
 \hline
 Point & $\lambda_{s'h'}$ & $\lambda_{s'}$ & $\lambda_{h'}$ & $v_s$ & $\tan(2\theta)$ \\ \hline
 A & $4.311\times 10^{-2}$ & $2.772\times 10^{-3}$ & 0.6858 & $2.462\times 10^4$ GeV & 36.996 \\
 B & $3.969\times 10^{-2}$ & $2.333\times 10^{-3}$ & 0.6936 & $2.474\times 10^4$ GeV & 10.46 \\
 C & $2.931\times 10^{-2}$ & $1.258\times 10^{-3}$ & 0.7171 & $2.597\times 10^4$ GeV & 3.048 \\
 D & $2.061\times 10^{-3}$ & $1.299\times 10^{-5}$ & 0.7778 & $7.816\times 10^4$ GeV & 0.335 \\
 E & $2.005\times 10^{-4}$ & $8.101\times 10^{-7}$ & 0.7819 & $2.472\times 10^5$ GeV & 0.101 \\
 F & $9.996\times 10^{-6}$ & $3.809\times 10^{-8}$ & 0.7823 & $1.106\times 10^6$ GeV & 0.0225 \\
\hline
\end{tabular}
\end{center}
\caption{Benchmark points.}
\label{results_table}
 \end{table*}

First, looking at $r_\epsilon$ and $\tan(2\theta)$ we notice that requiring maximal mixing implies slightly more fine-tuning than is needed for a nearly mass-degenerate spectrum with smaller mixing. If we let go of the requirement of maximal mixing, the amount of fine-tuning is then set by the mass-degeneracy $\delta$, and for our choise of $\delta=0.1$ this translates to fine-tuning of $r_\epsilon\sim\mathcal{O}(10^{-2})$, whereas $r_\epsilon\sim\mathcal{O}(10^{-4})$ is needed for maximal mixing.

Looking at the couplings $\lambda_i$ we notice a general pattern. The self-coupling of the messenger field $\lambda_s$ and the portal coupling $\lambda_{sh}$ are allways small, whereas the self-coupling of $h$ is larger. This structure is also apparent in equation (\ref{epsilon_definitions}). Notice that the normalization of $\lambda_h$ adopted in (\ref{scalar_potential}) differs from the usual normalization in the SM by a factor of four, $4\lambda_h=\lambda_h^{\rm SM}$, so that the values shown in table \ref{results_table} for $\lambda_h$ are in fact close to the SM value $\lambda_h^{\rm SM}\approx 0.13$. In terms of the physically measurable couplings $\lambda_{i'}$ of the mass-eigenstates the situation is similar. The self-coupling of one of the mass-eigenstates, $s'$, is much smaller than the self-coupling of the other state, which is order one. The coupling between the scalar fields $\lambda_{s'h'}$ is small, and vanishes in the limit of zero mixing.

The vacuum expectation value of the messenger field depends on the amount of mixing. It is $v_s\sim\mathcal{O}(10^4)$ GeV for maximal mixing, and grows large in the zero-mixing limit.

\section{Phenomenology of dark matter}
\label{DMsection}
As explained in the introduction, we imagine that the dark sector consists of a set of fields that are SM singlets but have internal quantum numbers and gauge interactions. The dark sector gauge interaction may be weak, as in a dark Coleman-Weinberg scenario or in the dark SUSY model; or it can be strong, as in the dark technicolor model in which case the dark matter candidate is a composite state. Here we will consider the DM field to be a singlet Majorana fermion. In general the DM sector can consist of fermions, scalars, or both, but for simplicity we will restrict ourselves to the fermionic case in this paper and postpone the fully general analysis until later.

The Lagrangian of the  DM field is
\be
 \mathcal{L}_\text{DM} =  \bar\psi (i \slashed \partial - M ) \psi + y s \bar\psi \psi ,
\ee
where $\psi$ is the DM Majorana fermion in Dirac notation, $M$ a tree-level mass term, whose origin is in the dynamics of the dark sector, and $y$ is the Yukawa coupling with the scalar singlet. We have omitted the gauge interaction of the fermion field, since we want to keep the analysis as general as possible and not restricted to a given model for the DM sector. The field $\psi$ may thus be an elementary fermion that gets a mass from the dynamics of the dark sector, or it may be a composite state whose mass originates from a strong interaction. We will not specify the nature of the DM field further and will treat the effective tree-level mass term $M$ as a free parameter.

\begin{figure}[t]
\begin{center}
\includegraphics[width=0.49\textwidth]{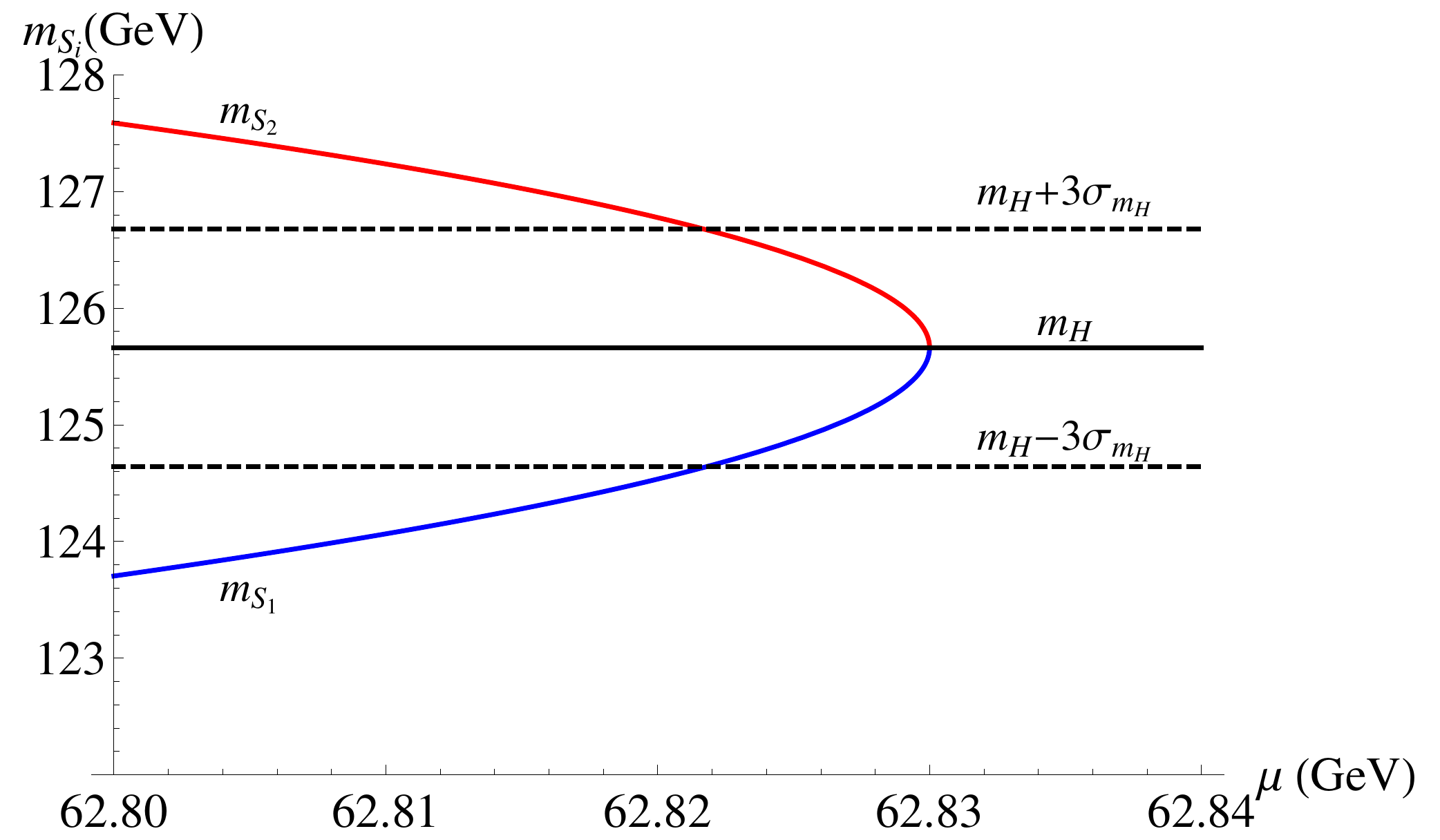}
\caption{$m_{S_{1,2}}$ in function of $\mu_s$.
The blue (red) line stands for $m_{S_1}$ $(m_{S_2})$. The black continuous line represents $m_H$ and the black dashed lines represent $m_H \pm 3 \sigma_{m_H}$.}
\label{fig:mSi}
\end{center}
\end{figure}

The direct detection cross section is given by \cite{Farina:2013mla}
\be
 \sigma_\text{SI} = \frac{y^2 \sin^2(2 \theta)}{8\pi} \frac{m^4_N f^2}{v_h^2} \left(\frac{1}{m_{S_1}^2}-\frac{1}{m_{S_2}^2}\right)^2, \label{sigmaSI}
\ee
where $m_N$ is the nucleon mass and $f$ is the nucleon matrix element.
This cross section is suppressed by the approximate degeneracy of the two scalar mass eigenstates. Note that this suppression is a general feature of this class of models, where the dark sector couples to the Higgs only via the messenger scalar, and is not specific only to the fermionic case that we have chosen for the excplicit calculation. Here we will focus on the scenario of maximal mixing, since this results in the most conservative limit for direct detection. If the mixing is smaller, the direct detection cross section is further suppressed by the smallness of the mixing angle, as is apparent in equation (\ref{sigmaSI}). Thus we will fix $\theta=\pi/4$, implying $\mu_s^2=\mu_h^2$ in terms of the parameters defined in equation (\ref{squaremassmatrix}). It is then natural to assume
\be
 \mu_s^2=\mu_h^2=m_H^2,
\ee
where $m_H$ is the central experimental Higgs mass value.
After these assumptions the only remaining free parameter in the scalar potential is $\mu$ and the mass eigenvalues are given by
\be
 m_{S_{1,2}}^2 = m_H^2 \pm m_H \sqrt{m_H^2-4 \mu^2}.
\ee
Moreover, $\mu$ is quite constrained by the uncertainty in the Higgs mass measurement. In Fig. \ref{fig:mSi} we plot the mass eigenvalues $m_{S_{1,2}}$ as a function of $\mu$.
The blue (red) line stands for $m_{S_1}$ $(m_{S_2})$. The black continuous line represents $m_H$ and the black dashed lines represent $m_H \pm 3 \sigma_{m_H}$.
We can see that $\mu \simeq m_H/2$ and that its maximum value is exactly $m_H/2$.

\begin{figure}[t]
\begin{center}
\includegraphics[width=0.45\textwidth]{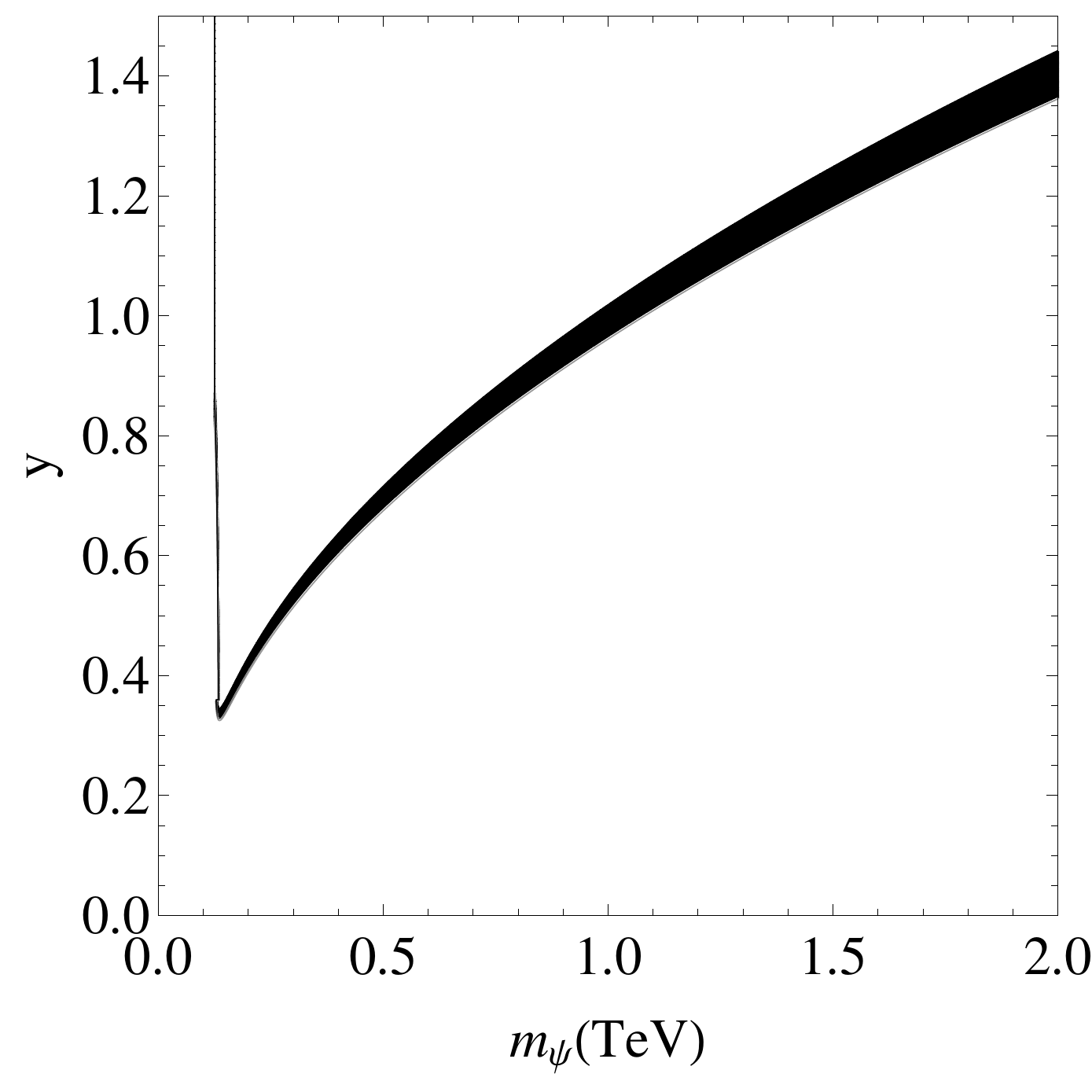}
\caption{Relic density estimation as function of $m_\psi$ and $y$.
The black region corresponds to a relic density in the range $\Omega_c h^2 \pm 5 \sigma$.}
\label{fig:relic}
\end{center}
\end{figure}

Before going back to the direct detection cross section, let us spend some words on the DM annihilation cross section.
DM can annihilate essentially via two types of processes:
\begin{itemize}
 \item[i.] $\psi \psi \to \varphi^\text{SM}_i \varphi^\text{SM}_j$
 \item[ii.] $\psi \psi \to S_i S_j$
\end{itemize}
where $\varphi^\text{SM}_{i,j}$ are SM particles (fermions or vectors).
The process i. is mediated by the exchange of virtual $S_{1,2}$ in the $s-$channel.
Therefore it is easy to check that it suffers the same suppression as $\sigma_\text{SI}$, so it can be neglected.
The process ii. is given by two types of interactions: the exchange of virtual $S_{1,2}$ in the $s-$channel, which is again negligible, and
the exchange of a virtual $\psi$ in the $t-$ and $u-$channels, which is the dominant process. The cross section can be obtained from equation (35) of \cite{Farina:2013mla}, by substituting $\cos\theta=\sin\theta=1/\sqrt{2}$. Since the mass splitting between $m_{S_{1,2}}$ does not play any relevant role in this computation, we use $m_{S_{1,2}}  \simeq m_H$ and $\mu \simeq m_H/2$ to obtain the formula
\bea
 \langle \sigma_{\psi \psi} v_\text{rel} \rangle &=& \sum_{i,i}  \langle \sigma_{\psi \psi \to S_i S_j} v_\text{rel} \rangle \\
  &\simeq&  v_\text{rel}^2
       \frac{y^4 m_\psi^2 \sqrt{1-\frac{m_H^2}{m_\psi^2}} \left(9 m_\psi^4-8 m_\psi^2 m_H^2+2 m_H^4\right)}{24 \pi \left(m_H^2-2 m_\psi^2\right)^4}, \nonumber
\eea
where $ m_\psi = |y v_s + M|$ is the DM mass and $v_\text{rel}$ is the relative velocity of the annihilating DM particles. 
The Planck Collaboration \cite{Ade:2013lta} measured the cold DM relic density to be $\Omega_c h^2 \pm \sigma = 0.1199 \pm 0.0027$.
We present our results in Fig. \ref{fig:relic} for the relic density estimation as a function of $m_\psi$ and $y$.
The black region corresponds to a relic density in the range $\Omega_c h^2 \pm 5 \sigma$, and the white region is for relic density outside of this range.

Since we have only two free parameters, the model is quite predictive.
Ignoring the experimental uncertainties, Fig. \ref{fig:relic} is the plot of the function $y_{\Omega h^2}(m_\psi)$,
obtained by solving the equation $\Omega_c h^2 = 0.1199$ as a function of $y$.
We can see that in order to obtain a correct relic density we need a Yukawa coupling $y$ roughly of the order of one. Thus, in order to get a fermion mass roughly of the order of 1 TeV, there must a fine tuned cancellation between the effective tree-level fermion mass $M$ and the contribution induced by $y v_s$, which is around 10 TeV or more. As the dominant annihilation process is to two scalars that in turn decay with the SM Higgs branching ratios dominantly to $b$-quarks and $W$-bosons, there will be no constraints from indirect detection of the annihilation \cite{Cirelli:2010xx}.

\begin{figure}[t!]
\begin{center}
\includegraphics[width=0.45\textwidth]{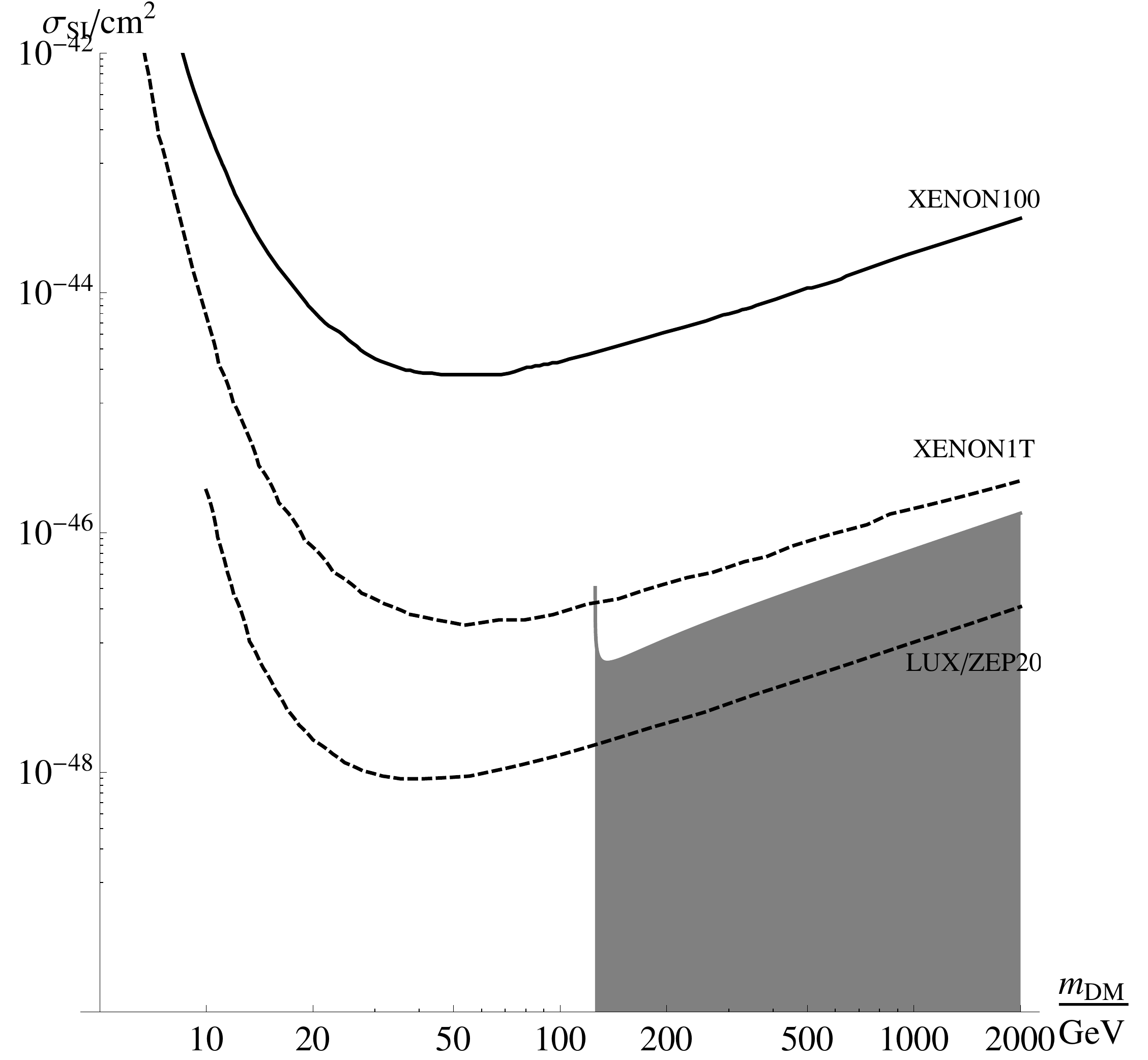}
\caption{The direct detection cross section region (in gray color) that corresponds to the allowed region by relic density measurements.
The black continuous line represents XENON100 bound for 2012, while the two black dashed lines stands for XENON1T and LUX/ZEP20 projections.}
\label{fig:SIplot}
\end{center}
\end{figure}

Now let us go back to the direct detection cross section, given in eq. (\ref{sigmaSI}).
In Fig. \ref{fig:SIplot} we plot in gray color the direct detection cross section region that corresponds to the allowed region by relic density measurements.
The black continuous line represents XENON100 bound for 2012 \cite{Aprile:2012nq},
while the two black dashed lines stands for XENON1T  \cite{Aprile:2012zx} and LUX/ZEP20  \cite{LUXZEP} projections\footnote{To produce the curves,
we used the online tool at http://dendera.berkeley.edu/plotter/entryform.html}.
Now of course the mass splitting between $S_{i,j}$ is relevant.
The maximum allowed splitting is clearly $6 \sigma_{m_H}$, which corresponds to the highest possible value for $\sigma_{SI}$,
represented by the border of the gray region in Fig. \ref{fig:SIplot}. Any other allowed mass splitting will decrease $\sigma_{SI}$.
We can see that we are always in agreement with the XENON100 bound, and in most of the parameter region DM will be undetectacble even for XENON1T,
but eventually detectable for LUX/ZEP20 unless the mass degeneracy is very extreme.
The only way to detect DM at XENON1T is that the DM mass is also very close to $m_H$.

\section{Discussion and Conclusions}

In light of the LHC data, namely the observation of the SM Higgs boson and the non-observation of any deviations from the SM or any new particles beyond the SM, a natural scenario for new physics is a hidden sector. This sector consists of SM singlets, but has internal dynamics that generate the DM mass scale. This scale is then mediated to the SM via a Higgs portal coupling, resulting in the vacuum expectation value for the Higgs and the usual electroweak symmetry breaking mechanism of the SM. We have studied this general setup in the region of the parameter space where the Higgs is nearly mass-degenerate and strongly mixed with the messenger scalar. In this case the observed Higgs signal is actually the sum of the two overlapping signals from the two nearly mass degenerate scalars.

We have shown that this scenario is plausible, although some finetuning is needed to achieve the highly degenerate mass spectrum. The direct detection cross section of the DM particle is suppressed by the mass degeneracy, and thus all existing bounds on direct detection are easily avoided.

\mysection{Acknowledgement}
This work was supported by the ESF grants 8499, 8943, MTT8, MTT59, MTT60, MJD140, MJD435, MJD298,
by the recurrent financing SF0690030s09 project and by the European Union through the European Regional Development Fund.


\begin{thebibliography}{99}

\bibitem{Aad:2012tfa}
  G.~Aad {\it et al.}  [ATLAS Collaboration],
  Phys.\ Lett.\ B {\bf 716}, 1 (2012)
  [arXiv:1207.7214 [hep-ex]].

\bibitem{Chatrchyan:2012ufa}
  S.~Chatrchyan {\it et al.}  [CMS Collaboration],
  Phys.\ Lett.\ B {\bf 716}, 30 (2012)
  [arXiv:1207.7235 [hep-ex]].

\bibitem{Giardino:2013bma}
  P.~P.~Giardino, K.~Kannike, I.~Masina, M.~Raidal, A.~Strumia,
  arXiv:1303.3570 [hep-ph].

\bibitem{Ellis:2013lra}
  J.~Ellis, T.~You,
  arXiv:1303.3879 [hep-ph].

\bibitem{Djouadi:2013qya}
  A.~Djouadi, G.~使.~Moreau,
  arXiv:1303.6591 [hep-ph].

\bibitem{Falkowski:2013dza}
  A.~Falkowski, F.~Riva, A.~Urbano,
  arXiv:1303.1812 [hep-ph].

\bibitem{Ade:2013lta}
  P.~A.~R.~Ade {\it et al.}  [ Planck Collaboration],
  arXiv:1303.5076 [astro-ph.CO].

\bibitem{Bertone:2004pz} 
For reviews see,   G.~Bertone, D.~Hooper and J.~Silk,
  Phys.\ Rept.\  {\bf 405}, 279 (2005)
  [hep-ph/0404175].
  L.~Bergstrom,
  Annalen Phys.\  {\bf 524}, 479 (2012)
  [arXiv:1205.4882 [astro-ph.HE]].



\bibitem{Hempfling:1996ht}
  R.~Hempfling,
  Phys.\ Lett.\ B {\bf 379}, 153 (1996)
  [hep-ph/9604278];
  K.~A.~Meissner and H.~Nicolai,
  Phys.\ Lett.\ B {\bf 648}, 312 (2007)
  [hep-th/0612165];
  W.~-F.~Chang, J.~N.~Ng and J.~M.~S.~Wu,
  Phys.\ Rev.\ D {\bf 75}, 115016 (2007)
  [hep-ph/0701254 [HEP-PH]];
  R.~Foot, A.~Kobakhidze, K.~.L.~McDonald and R.~.R.~Volkas,
  Phys.\ Rev.\ D {\bf 76}, 075014 (2007)
  [arXiv:0706.1829 [hep-ph]];
  R.~Foot, A.~Kobakhidze and R.~R.~Volkas,
  Phys.\ Lett.\ B {\bf 655}, 156 (2007)
  [arXiv:0704.1165 [hep-ph]];
  R.~Foot, A.~Kobakhidze, K.~L.~McDonald and R.~R.~Volkas,
  Phys.\ Rev.\ D {\bf 77}, 035006 (2008)
  [arXiv:0709.2750 [hep-ph]];
  R.~Foot, A.~Kobakhidze and R.~R.~Volkas,
  Phys.\ Rev.\ D {\bf 82}, 035005 (2010)
  [arXiv:1006.0131 [hep-ph]];
  L.~Alexander-Nunneley and A.~Pilaftsis,
  JHEP {\bf 1009}, 021 (2010)
  [arXiv:1006.5916 [hep-ph]].
  C.~Englert, J.~Jaeckel, V.~V.~Khoze and M.~Spannowsky,
  JHEP {\bf 1304}, 060 (2013)
  [arXiv:1301.4224 [hep-ph]].

\bibitem{Hur:2011sv}
  T.~Hur and P.~Ko,
  Phys.\ Rev.\ Lett.\  {\bf 106}, 141802 (2011)
  [arXiv:1103.2571 [hep-ph]].

\bibitem{Heikinheimo:2013fta}
  M.~Heikinheimo, A.~Racioppi, M.~Raidal, C.~Spethmann and K.~Tuominen,
  arXiv:1304.7006 [hep-ph].

\bibitem{Heikinheimo:2013xua}
  M.~Heikinheimo, A.~Racioppi, M.~Raidal, C.~Spethmann and K.~Tuominen,
  arXiv:1305.4182 [hep-ph].

\bibitem{Farina:2013mla} 
  M.~Farina, D.~Pappadopulo and A.~Strumia,
  arXiv:1303.7244 [hep-ph].

\bibitem{Dubovsky:2013ira} 
  S.~Dubovsky, V.~Gorbenko and M.~Mirbabayi,
  arXiv:1305.6939 [hep-th].

\bibitem{CMS2Higgs}
  CMS collaboration,
  CMS-PAS-HIG-13-016.


\bibitem{Gunion:2012gc}
  J.~F.~Gunion, Y.~Jiang and S.~Kraml,
  Phys.\ Rev.\ D {\bf 86}, 071702 (2012)
  [arXiv:1207.1545 [hep-ph]].
  J.~F.~Gunion, Y.~Jiang and S.~Kraml,
  Phys.\ Rev.\ Lett.\  {\bf 110}, 051801 (2013)
  [arXiv:1208.1817 [hep-ph]].
  P.~M.~Ferreira, R.~Santos, H.~E.~Haber and J.~P.~Silva,
  Phys.\ Rev.\ D {\bf 87}, 055009 (2013)
  [arXiv:1211.3131 [hep-ph]].
  A.~Drozd, B.~Grzadkowski, J.~F.~Gunion and Y.~Jiang,
  JHEP {\bf 1305}, 072 (2013)
  [arXiv:1211.3580 [hep-ph]].
  Y.~Grossman, Z.~'e.~Surujon and J.~Zupan,
  JHEP {\bf 1303}, 176 (2013)
  [arXiv:1301.0328 [hep-ph]].
  A.~Efrati, D.~Grossman and Y.~Hochberg,
  arXiv:1302.7215 [hep-ph].

\bibitem{ScalarMixing}
  B.~Patt and F.~Wilczek,
  hep-ph/0605188.
  C.~Englert, T.~Plehn, D.~Zerwas and P.~M.~Zerwas,
  Phys.\ Lett.\ B {\bf 703}, 298 (2011)
  [arXiv:1106.3097 [hep-ph]].
  R.~S.~Gupta and J.~D.~Wells,
  Phys.\ Lett.\ B {\bf 710}, 154 (2012)
  [arXiv:1110.0824 [hep-ph]].
  D.~Bertolini and M.~McCullough,
  JHEP {\bf 1212}, 118 (2012)
  [arXiv:1207.4209 [hep-ph]].
  B.~Batell, D.~McKeen and M.~Pospelov,
  JHEP {\bf 1210}, 104 (2012)
  [arXiv:1207.6252 [hep-ph]].
  M.~Fairbairn and R.~Hogan,
  arXiv:1305.3452 [hep-ph].
  S.~Chpoi, S.~Jung and P.~Ko,
  arXiv:1307.3948 [hep-ph].

\bibitem{Cirelli:2010xx}
  M.~Cirelli, G.~Corcella, A.~Hektor, G.~Hutsi, M.~Kadastik, P.~Panci, M.~Raidal and F.~Sala {\it et al.},
  JCAP {\bf 1103} (2011) 051
   [Erratum-ibid.\  {\bf 1210} (2012) E01]
  [arXiv:1012.4515 [hep-ph]].

\bibitem{Aprile:2012nq}
  E.~Aprile {\it et al.}  [XENON100 Collaboration],
  Phys.\ Rev.\ Lett.\  {\bf 109}, 181301 (2012)
  [arXiv:1207.5988 [astro-ph.CO]].

\bibitem{Aprile:2012zx}
  E.~Aprile [XENON1T Collaboration],
  arXiv:1206.6288 [astro-ph.IM].

\bibitem{LUXZEP}
LUX/ZEP III Collaboration, July 2008, Gaitskell, Brown University





\end{thebibliography}
\end{document}